# Viscoelastic Properties of Vascular Endothelial Cells Exposed to Stretch

Kathryn Osterday[a)*], Thomas Chew[**], Phillip Loury[**], Jason Haga[**], Manuel Gómez-González[*], Juan C. del Álamo[*] & Shu Chien[***]

[*]*Department of Mechanics and Aerospace Engineering, University of California San Diego (UCSD), San Diego, 92092, California USA*
[**]*Department of Bioengineering, UCSD, San Diego, 92092, California USA*
[***]*Department of Bioengineering and Medicine, UCSD, San Diego, 92092, California USA*

In this paper, we study how cytoskeletal remodeling is correlated to changes in subcellular microrheology. We analyze the changes in the magnitude and directionality of the shear and elastic moduli of bovine aortic endothelial cells (BAECs) exposed to cyclical, uniaxial stretch. We find that, when stretched, BAECs stiffen and align their softest direction of mechanical polarization perpendicular to stretch. We hypothesize that the response of VECs to stretch acts to minimize intracellular strain in response to stress.

## INTRODUCTION

Vascular endothelial cells (VECs) line the lumen of blood vessels. Apart from serving as an impermeable barrier, VECs produce, secrete, and metabolize biochemical substances that are critical to cardiovascular function. The Response of VECs to mechanical forces plays a significant role in regulating vascular performance. Atherosclerosis has a propensity for arterial branch points and local lumen expansions - regions where the VECs are subjected to disturbed flow conditions consisting of flow separation, reversal and reattachment. These regions experience biaxial stretch, low shear stress magnitude, high shear stress gradient and little net direction of flow. In contrast, the straight regions of blood vessels, which are exposed to uniaxial stretch, high shear stress and directional flow, are spared from atherogenesis.

The pulsatile nature of blood flow in arteries exposes VECs to primarily two types of mechanical stresses - velocity gradients expose VECs to shear stress, whereas the pressure pulse expands the artery and stretches the lumen where VECs are seeded. Widely studied, shear stress is known to soften VECs and to align VECs parallel to the direction of shear [2]. Previous work shows that uniaxial stretch aligns VECs perpendicular to the direction of stretch but the effects of uniaxial stretch on the mechanical properties of the cytoplasm have not been extensively studied.

In vivo, both shear and stretch have the same net effect on actin cytoskeletal alignment, for they align the cytoskeleton parallel to the direction of blood flow. Cytoskeletal remodeling is correlated to changes in subcellular microrheology, which regulates the mechanotransduction process by redistributing the external stresses across subcellular domains. This process is partially responsible for a wide variety of cardiovascular functions, some of which include intracellular signaling, gene expression, smooth muscle cell contraction, cell remodeling, and the secretion and metabolism of biochemical substances. Both stretch and shear may act in tandem to maintain cardiovascular homeostasis in order to minimize intracellular strain in response to stress.

## MATERIALS AND METHODS

**Cell culture and stretch chamber preparation**
We subject bovine aortic endothelial cells (BAECs) to periodic cycles of uni-axial stretch using a device that allows *in situ* live cell imaging [1].

**Removing Stage Drift**
BAECs are seeded on a 127-micron thick silicone membrane; this diminishes the quality of the acquired videos. Because our particle tracking method relies on subpixel resolution of particle position, the videos are carefully processed to remove stage drift. Our stage-motion detection algorithm optimizes the correlation between two subsequent images by varying the translation of the second image. To translate images, we employ a cubic-spline interpolation algorithm. As an initial guess for the translation coordinates, we apply phase correlation to a manually-cropped, rectangular region of both images. To ensure the algorithm is robust regardless of whether cells change significantly over the length of the video, we compare subsequent frames and dedrift each frame by the sum of the motion of all previous frames.

**Directional Particle Tracking Microrheology (DPTM)**
We use DPTM to calculate the rheological properties of the cytoplasm along the principal directions of mechanical polarization. Our method is an improvement of the technique previously described by del Alamo *et al* (2008) [2]. Our custom-written MATLAB functions track the motion of intracellular particles, mitochondria specifically, and calculate the principle values of the 2D mobility tensor, to obtain the MSDs along the stiffest and softest directions of the cell; to which we apply novel microrheology formulae developed by Gomez-Gonzalez (2012) to calculate the viscoelastic shear moduli from these MSDs. These equations assume a constitutive equation that models the cytoplasm as an orthotropic fluid. The new formulae decouple the effect of the viscosity in each orthogonal direction as the particle moves along each direction of mechanical polarization [3]. The shear and elastic moduli are numerically calculated by setting the response function given by Einstein's equation for Brownian motion equal to the response functions for the orthotropic medium.

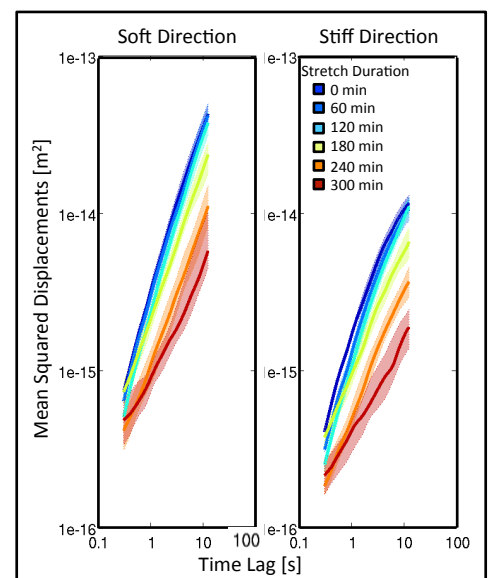

Figure 1

## RESULTS

At time separations less than 10 seconds, the value of the mean squared displacements (MSDs) were on average two to three times larger along the soft direction than along the stiff one (*Figure 1*). The ratios of the MSDs along each direction give an indication of the degree of anisotropy. The level of anisotropy of the MSDs did not change significantly after stretch.

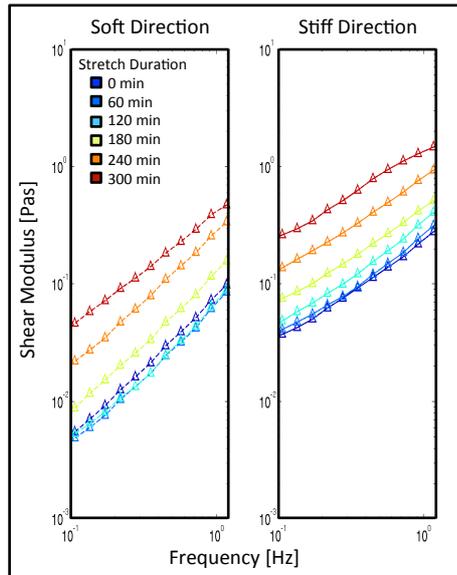

Figure 2

The values of the MSDs decreased significantly after stretch (*Figure 1*). The value of the shear modulus, calculated using the slope of the MSDs, increased accordingly (*Figure 2*). The logarithmic slopes of the MSDs with respect to time separation also varied over the course of the experiment. These slopes reveal how close the viscoelastic behavior of the cytoplasm is to being liquid-like (unity slope) or elastic-like (zero slope). The slopes remained fairly constant for the first two hours of stretch with values close to 1, and decreased approximately 10% after each subsequent hour of stretch. This decrease in slope indicates that VECs become less liquid and more elastic under stretch.

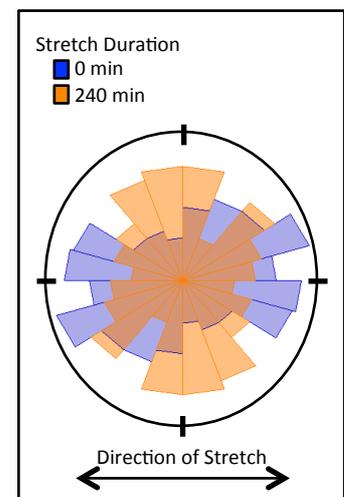

Figure 3

*Figure 3* shows the distribution of the orientation of maximum compliance for a time separation of 10 seconds. The cells aligned their softest direction perpendicular to stretch after 240 min.

## CONCLUSIONS

The pressure pulse present in arteries exposes VECs to stretch. Under stretch, we have shown that VECs become more elastic-like, stiffen and align their softest direction of mechanical polarization perpendicular to the direction of stretch. This response likely acts to minimize intracellular strain. Both shear and stretch align the softest direction of mechanical polarization parallel to the direction of blood flow in vivo. It is widely believed that this orientation helps the cell sense the direction and magnitude of external stimuli.